\documentclass[prd,twocolumn,nofootinbib,aps,tightenlines,preprintnumbers,notitlepage,longbibliography,superscriptaddress]{revtex4-1}
\usepackage{graphicx}
\usepackage{amsmath}
\usepackage{amsfonts}
\usepackage[colorlinks=true,citecolor=blue,urlcolor=cyan,linkcolor=black]{hyperref}
\usepackage{relsize}
\usepackage{pgfplots}

\pgfplotsset{compat=1.9}

\definecolor{indigo}{HTML}{4B0082}
\definecolor{purple}{HTML}{800080}
\definecolor{firebrick}{HTML}{b22222}
\definecolor{salmon}{HTML}{fa8072}
\definecolor{orange}{HTML}{FFA500}
\definecolor{blue}{HTML}{1E90FF}

\usepackage{graphicx}
\usepackage{dcolumn}
\usepackage{bm}
\usepackage[normalem]{ulem}
\usepackage{xcolor}
\usepackage{subfigure}

\usepackage{footnote}

\graphicspath{ {figures/} }

\usepackage{epstopdf}
\newcommand{\be}{\begin{eqnarray}}
\newcommand{\non}{\nonumber \\}
\newcommand{\ee}{\end{eqnarray}}

\usepackage[export]{adjustbox}

\usepackage{float}

\usepackage{suffix}
\usepackage{mathtools}

\DeclarePairedDelimiterX\MeijerM[3]{\lparen}{\rparen}%
{\begin{smallmatrix}#1 \\ #2\end{smallmatrix}\delimsize\vert\,#3}

\newcommand\MeijerG[8][]{%
  G^{\,#2,#3}_{#4,#5}\MeijerM[#1]{#6}{#7}{#8}}

\WithSuffix\newcommand\MeijerG*[7]{%
  G^{\,#1,#2}_{#3,#4}\MeijerM*{#5}{#6}{#7}}

\usepackage{hyperref}
\hypersetup{
    bookmarks=true,         
    unicode=false,          
    pdftoolbar=true,        
    pdfmenubar=true,        
    pdffitwindow=false,     
    pdfstartview={FitH},    
    pdftitle={My title},    
    pdfauthor={Author},     
    pdfsubject={Subject},   
    pdfcreator={Creator},   
    pdfproducer={Producer}, 
    pdfkeywords={keyword1, key2, key3}, 
    pdfnewwindow=true,      
    colorlinks=true,       
    linkcolor=black,          
    citecolor=black,        
    filecolor=magenta,      
    urlcolor=cyan           
}

\newcommand{\aap}{Astron. Astrophys.}

\newcommand{\aj}{Astron. J.}

\newcommand{\mnras}{Mon. Not. R. Astron. Soc.}
\newcommand{\jcap}{J. Cosm. Astropart. Phys.}

\newcommand{\physrep}{Phys. Rep.}

\newcommand{\dd}{{\rm d}}

\begin{document}

\title{Probing ultra-light axions with the 21-cm Signal during Cosmic Dawn}

\newcommand{\jhu}{Department of Physics \& Astronomy, Johns Hopkins University, Baltimore, MD 21218, USA}

\newcommand{\kcl}{Theortetical Particle Physics and Cosmology, King's College London, Strand, London, WC2R 2LS, United Kingdom}

\author{Selim~C.~Hotinli}
\email{shotinl1@jh.edu}
\affiliation{\jhu}

\author{David~J.~E.~Marsh}
\email{david.j.marsh@kcl.ac.uk}
\affiliation{\kcl}

\author{Marc~Kamionkowski}
\email{kamion@jh.edu}
\affiliation{\jhu}

\date{\today}

\begin{abstract}
Ultra-light axions (ULAs) are a promising and intriguing set of dark-matter candidates.  We study the prospects to use forthcoming measurements of 21-cm fluctuations from cosmic dawn to probe ULAs.  We focus in particular on the velocity acoustic oscillations (VAOs) in the large-scale 21-cm power spectrum, features imprinted by the long-wavelength ($k\sim0.1\,{\rm Mpc}^{-1}$) modulation, by dark-matter--baryon relative velocities, of the small-scale ($k\sim 10-10^3\, {\rm Mpc}^{-1}$) power required to produce the stars that heat the neutral hydrogen.  Damping of small-scale power by ULAs reduces the star-formation rate at cosmic dawn which then leads to a reduced VAO amplitude. Accounting for different assumptions for feedback and foregrounds, experiments like HERA may be sensitive to ULAs with masses up to $m_{\alpha}\approx 10^{-18}\text{eV}$, two decades of mass higher than current constraints.
\end{abstract}

\preprint{KCL-PH-TH/2021-94}

\maketitle

\section{Introduction}

 Next-generation cosmic microwave background (CMB) experiments such as the Simons Observatory (SO)~\citep{Ade:2018sbj,Abitbol:2019nhf} and CMB-S4~\citep{Abazajian:2016yjj}, galaxy surveys such as DESI~\citep{Aghamousa:2016zmz} and the Vera Rubin Observatory (VRO)~\citep{2009arXiv0912.0201L}, as well as 21-cm  experiments such as HERA~\citep{DeBoer:2016tnn} or SKA1-low~\cite{2019arXiv191212699B,Bacon:2018dui} will soon generate a wealth of stringent new tests of the standard cosmological model and hopefully shed light on the nature of dark matter.  
 Here, we consider ultra-light axions (ULAs), a dark-matter candidate that is compelling because of its possible connections to the strong-CP problem, galactic substructure, and string theory~\cite{Arvanitaki:2009fg,Marsh:2015xka}. 

Here we explore new probes of ULAs enabled by forthcoming measurements of fluctuations of the 21-cm line of neutral hydrogen  during cosmic dawn.  The galaxy-formation rate in any particular region of the Universe can be affected by the dark-matter--baryon relative velocity in that region.  This thus introduces a fluctuation in galaxy-formation rate on the fairly large ($k\sim 0.1\,h$Mpc$^{-1}$) coherence scale of the linear-theory dark-matter--baryon relative velocity \cite{Tseliakhovich:2010bj}.  The oscillations in the dark-matter--baryon relative-velocity power spectrum are thus imprinted in the power spectrum of 21-cm fluctuations from cosmic dawn \cite{Dalal:2010yt,Naoz:2011if,Greif:2011iv,McQuinn:2012rt,Stacy:2010gg,Naoz:2011if,Fialkov:2011iw,Yoo:2011tq,Pritchard:2011xb,Ali-Haimoud:2013hpa,Barkana:2016nyr,Munoz:2019rhi,Munoz:2019fkt,Munoz:2019hjh,Hotinli:2021xln}.

ULAs (and other models of ultralight bosonic dark mater) differ from standard cold dark matter (CDM) due to their astrophysically large de Broglie wavelength, which leads to an effective sound speed in the ULA density perturbations, suppressing the growth of cosmic structure relative to CDM~\cite{Khlopov:1985jw,Arvanitaki:2009fg,Marsh:2010wq}.\footnote{{Note that recent studies of ULAs using $N$-body simulations suggest interference effects which can lead to $\sim10-15\%$ effect on the matter power-spectrum on scales comparable to the de Broglie wavelength of the fluid~(see e.g. Refs.~\citep{Li:2018kyk,Zhang:2018ghp,Veltmaat:2019hou,Schwabe:2021jne}), which we will omit in this paper. The intereference and vorticity effects that are not captured by the fluid formulation we use here are unlikely to be significant at the power spectrum level, although a possible effect on early star formation is still an open question~\citep{Nori:2018pka,Mocz:2019pyf}}.} This effect, and other effects related to the ULA wavelength, can be probed by a wide variety of astrophysical and cosmological measurements~\cite{Amendola:2005ad,Arvanitaki:2009fg,Marsh:2015xka,Hui:2016ltb,Marsh:2021lqg}. The CMB and galaxy surveys establish a lower bound on the ULA mass around $10^{-25}\,\text{eV}$, and probe the ULA density fraction at the percent level for lower masses~\cite{Hlozek:2014lca,Hlozek:2017zzf,Lague:2021frh}. The lower limit can be improved by a variety of cosmological probes (e.g. Refs.~\cite{Bozek:2014uqa,Schive:2015kza,Corasaniti:2016epp,Dentler:2021zij}), with the strongest, $m>2\times 10^{-20}\,\text{eV}$, being provided by the Lyman-alpha forest flux power spectrum \cite{Rogers:2020ltq}. 

The prospects to use the 21-cm signal as probe the small-scale matter power spectrum during cosmic dawn were discussed in Ref.~\citep{Munoz:2019hjh}.  Here we show that the velocity acoustic oscillations (VAOs) \cite{Munoz:2019fkt} in the 21-cm power spectrum can extend the sensitivity to an ULA mass up to two orders of magnitude higher.  The probe is quite robust to theoretical uncertainties as it relies on the linear-theory effects of ULAs, which are better understood than nonlinear effects.  Ref.~\cite{Marsh:2015daa} showed that if the ULA Jeans scale is larger than the baryon Jeans scale, then the modulation feature in the power spectrum caused by dark-matter--baryon relative motion is absent, thus suggesting that the VAO feature would also be absent. Ref.~\cite{Marsh:2015daa} thus speculated that a future measurement of VAOs would be able to probe ULA particle masses around $10^{-18}\,\text{eV}$. The present work addresses this claim in more detail. We compute explicitly the 21-cm cosmic dawn VAO feature from simulations and calculate the sensitivity of its amplitude to the ULA Jeans scale following Refs.~\citep{Marsh:2015daa,2010JCAP...11..007D,Dalal:2010yt,Ali-Haimoud:2013hpa,Munoz:2018jwq,Munoz:2019rhi}. We forecast the sensitivity of a realistic 21-cm measurement to this effect, accounting for foreground removal and baryonic feedback. 

This paper is organised as follows. We describe the 21-cm hydrogen line and relative-velocity signature in Section~\ref{sec:21cm_ps}. We discuss the ULAs and their effects on observables in Section~\ref{sec:ULAs}. We provide forecasts on the ULA mass in Section~\ref{sec:forecasts} and conclude with discussion in Section~\ref{sec:discussion}.

\section{The impact of relative velocities on the 21-cm power spectrum}\label{sec:21cm_ps}

\subsection{The 21-cm hydrogen line}

The absorption/emission of 21-cm photons by neutral hydrogen is determined by the $T_s$ of neutral hydrogen, which can be obtained from,
\be
\frac{n_1}{n_0}=\frac{g_1}{g_0}e^{-T_*/T_s}.
\ee
Here, $n_0$ and $n_1$ are the comoving number densities of the hydrogen atoms in the singlet  and the triplet states, where $g_0=1$  and $g_1=3$ are their numbers of degrees of freedom, respectively. The temperature corresponding to the 21-cm hyperfine transition is $T_*=0.068$\,K.  
The hydrogen emits (absorbs) photons from the CMB when the local spin temperature is higher (lower) than the CMB temperature. The distribution of these photons at the different wavelengths and at different redshifts can be used to infer the astrophysical and cosmological properties of our Universe. 

Our observable is the 21-cm brightness temperature~\citep{2006PhR...433..181F},
\be\label{eq:21cm_temperature}
T_{21} = 38{\rm mK}\, \left(1-\frac{T_\gamma}{T_s}\right)\left(\frac{1+z}{20}\right)^{\!1/2}\!\!x_{\rm HI}(1+\delta_b)\frac{\partial_rv_r}{H(z)}\,,\non
\ee
where $x_{\rm HI}$ is the neutral-hydrogen fraction, $\partial_r v_r$ is the line-of-sight gradient of the velocity, $T_\gamma$ is the CMB temperature, $\delta_b$ is the baryon overdensity, and $H$ is the Hubble parameter \citep{Barkana:2016nyr,Pritchard:2011xb,2006PhR...433..181F}.

We define the cosmic-dawn era to be that when the first stars formed, starting around $z \sim 25-35$~\citep{2014MNRAS.437L..36F}.  
After recombination, the gas kinetic temperature is dominated by adiabatic cooling and the spin temperature is coupled to the CMB temperature due to the high gas density and through collisions~\citep{Loeb:2003ya}.
Towards the the end of the dark ages, collisional coupling of hydrogen becomes ineffective and the 21-cm signal becomes small. The first stars produce UV photons that redshift into the Lyman-$\alpha$ line and couple the kinetic and spin temperatures of hydrogen in the intergalactic medium (IGM) via the Wouthuysen-Field effect \citep{1952AJ.....57R..31W, 1958PIRE...46..240F,Hirata:2005mz}.
Remnants of these first stars likely produce a background of $\sim 0.1 - 2 ~\rm keV$ $X$-rays~\citep{Pacucci:2014wwa,Pritchard:2006sq} heating the IGM before reionization progresses largely after $z \sim 10$~\citep{Morales:2009gs,Mesinger:2018ndr,Wise:2019qtq}. Toward the end of reionization ($z\lesssim 10$), Lyman-Werner feedback reduces the effects of streaming velocities in the IGM on the 21-cm signal\citep{2014MNRAS.437L..36F,FGU}.

We model the 21-cm fluctuations during cosmic-dawn with semi-numerical simulations provided by \texttt{21cmvFAST}\footnote{\href{https://github.com/JulianBMunoz/21cmvFAST}{github.com/JulianBMunoz/21cmvFAST}}, which is built upon  \texttt{21cmFAST}\footnote{\href{https://github.com/andreimesinger/21cmFAST}{github.com/andreimesinger/21cmFAST}}. Initial conditions for peculiar velocity and density fields are set at $z=300$ with a Gaussian random field in Lagrangian space, before being evolved with the Zel'dovich approximation \citep{1970A&A.....5...84Z} to match the mean collapse fraction for the conditional Sheth-Tormen halo mass function \citep{1999MNRAS.308..119S}.
The photon emission rates of the sources embedded in each halo are assumed proportional to the increase of the total collapsed halo mass. 
The excursion set formalism is used in each cell to estimate the mean number of sources contributing to the gas temperature from the surroundings.
The kinetic temperature is calculated including adiabatic expansion, Compton scattering with the CMB~\citep{2011ascl.soft06026S}, and the inhomogeneous heating history of the IGM (please see for details on this calculation in Refs.~\citep{2011MNRAS.411..955M,Murray:2020trn}).

For each observed frequency, we produce realizations of the 21-cm signal of coeval cubes of size 2000 Mpc on 2000$^3$ grids. Each coeval cube is simulated at the respective redshift from an initial density field given by appropriate transfer functions for relative velocities and matter. The power spectrum $\Delta_{21}^2(k)$ of 21-cm fluctuations satisfy
\be
\langle \delta T_{21}(\vec{k},z) \delta T_{21}^*(\vec{k}',z) \rangle = (2\pi)^3\delta(\vec{k}-\vec{k}') \frac{2\pi^2}{k^3} \Delta_{21}^2(k,z)\,,\ \ \ \
\ee
where $\delta T_{21}(\vec{k},z)$ is the Fourier transform of $[ T_{21}(\vec{x}) - \bar{T}_{21} ] / \bar{T}_{21}$, the zero-mean fluctuations of the 21-cm brightness temperature at redshift $z$. 

\subsection{The impact of relative velocities}\label{sec:impact_of_v}

\begin{figure}[b]
    \vspace{-0.25cm}
    \includegraphics[width=\columnwidth]{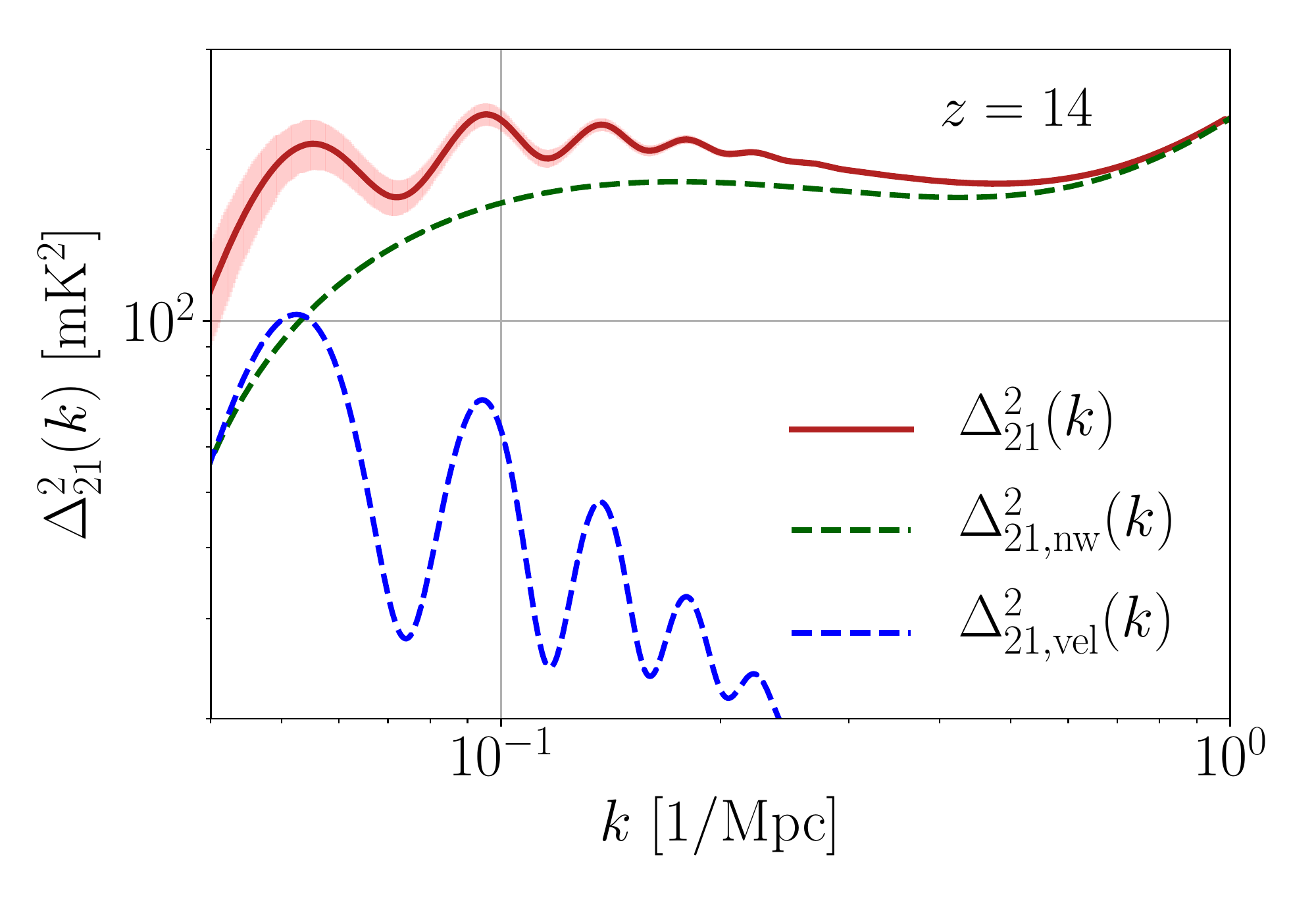}
    \vspace{-0.75cm}
    \caption{The brightness-temperature power spectra of t he 21-cm hydrogen line at redshift $z=14$ (in solid red), shown with fitted contribution to the power spectra (in dashed green) in the absence of the relative velocity effect and the velocity contribution (dashed blue) as discussed in Section~\ref{sec:impact_of_v}. For the simulations we used \texttt{21cmvFAST}~\citep{Munoz:2019fkt} with medium feedback. Error bars shown in the figure for the signal are Poisson errors from our simulations. The dashed green line is a fourth-order polynomial fit, as discussed in Section~\ref{sec:impact_of_v}.}
    \label{fig:display_CIPs_VAOs}
\end{figure}
\begin{figure*}[t!]
    \vspace{-0.25cm}
    \includegraphics[width=2\columnwidth]{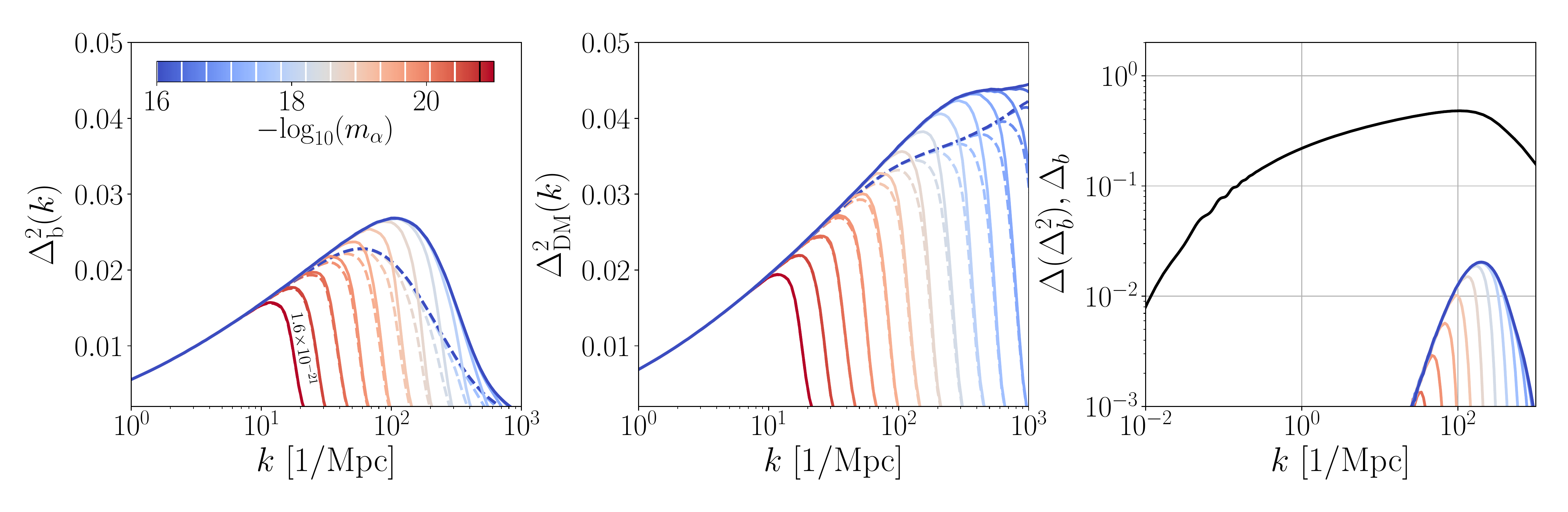}
    \vspace{-0.35cm}
    \caption{\textbf{The effect of ULAs on the baryon and DM power-spectra.} \textit{The left and middle plots} show the baryon and dark matter power-spectra for varying ULA mass, respectively. The ULAs can be observed to dampen the structure formation on small scales. Here, the solid (dashed) colored lines correspond to the power spectra in the absence (presence) of the relative velocity effect as shown in e.g.~Refs.~\citep{Dalal:2010yt,2010JCAP...11..007D,Ali-Haimoud:2013hpa,Munoz:2018jwq}. \textit{The right plot} compares the  characteristic baryon overdensity {$\Delta_b(k)=[\int_0^k k'^3\dd k'/(2\pi^2)P_b(k')]^{1/2}$} (solid black line) in the absence of ULAs with the characteristic change in the small-scale baryon power  about their mean value $\Delta(\Delta_b^2) \equiv |\Delta_b^2(k,v_{\rm cb}=0)-\Delta_b^2(k,v_{\rm cb})|$ (colored lines) as function of wavenumber $k$. The amplitude of the quadratic small-scale fluctuations ($k\sim100h\text{Mpc}^{-1}$) shown as comparable with the linear power-spectra on large scales ($k\sim0.01h\text{Mpc}^{-2}$). \textit{On all plots}, different colors correspond to the scenarios with ULAs with varying mass at redshift $z=20$. {The masses are highlighted with white markers on the colorbar shown on the left panel. The right-most line (highlighted with the black marker on the colorbar) corresponds to $m_\alpha=1.6\times10^{-21}{\rm eV}$, with succeeding lines to the left corresponding to masses as highlighted on the colorbar.}}
    \vspace*{-0.2cm}
    \label{fig:pedagogy_1}
\end{figure*}
The modulation induced by the dark-matter--baryon relative velocities on the 21-cm power spectrum can be captured from the statistics of the collapsed baryonic density to a good approximation. 
Briefly, the effect of bulk relative velocities is analogous to that of the gas pressure, which suppresses the accretion of baryons. The bulk kinetic energy of the gas gets converted into thermal energy as it falls into the dark matter (DM) halo, resulting in a change in the effective sound speed $c_{\rm eff,s}\simeq(c_s^2+v_{\rm cb}^2)^{1/2}$, hence in the baryon collapsed fraction~\citep{Dalal:2010yt,Naoz:2011if,Tseliakhovich:2010yw,Greif:2011iv,McQuinn:2012rt,Stacy:2010gg,Naoz:2011if,Fialkov:2011iw,Yoo:2011tq,Pritchard:2011xb,Barkana:2016nyr,Munoz:2019rhi}.
The effect of the relative velocities on the 21-cm brightness temperature power spectrum amplitude can then be parameterised as~\citep{Munoz:2019fkt}
\be\label{eq:velo_pert}
\Delta_{21,\rm vel}^2(k,z)=A_{\rm vel}(z)\Delta_{v^2}^2(k,z)|W(k,z)|^2\,,
\ee
where $A_{\rm vel}$ is some redshift-dependent amplitude of fluctuations.
The window function $W(k,z)$ depends on the various contributors to the 21-cm power spectrum such as Lyman-$\alpha$ coupling and $X$-ray heating.  Here, we defined $\Delta_{v^2}^2(k)$ as the power spectrum of the quantity 
\be
\delta_{v^2}=\sqrt{\frac{3}{2}}\left(\frac{v_{\rm cb}^2}{v_{\rm rms}^2}-1\right)\,,
\ee
which accurately captures the shape of the effect of relative velocities on the observables for the scales where the `streaming' bulk relative velocity can be approximated with a root-mean-squared value $v_{\rm rms}\simeq30 \,{\rm km\,s}^{-1}$ at recombination~\citep{Tseliakhovich:2010bj}. As the VAOs are statistically independent from the density fluctuations at first order, the amplitude of the 21-cm power spectrum can be written as
\be\label{eq:vao_definition}
\Delta_{21}^2(k,z)=\Delta_{21,\rm vel}^2(k,z)+\Delta^2_{21,\rm nw}(k,z)\,,
\ee
where $\Delta^2_{21\!,\rm nw}(k,z)$ is the component \textit{excluding} VAOs. Throughout this paper we use a fourth-order polynomial to parametrise the smooth contribution to the spectra following Ref.~\cite{Munoz:2019rhi},
\be\label{eq:smooth_pw}
\ln[\Delta_{21,\rm nw}^2(k,z)]=\sum_{i=0}^{4}c_i(z)[\ln k]^i\,,
\ee
where $c_i(z)$ are coefficients we fit for using simulations we discussed above. The fitted smooth spectra serve as our phenomenological model that captures the dependence of the 21-cm power spectra to the cosmological parameters. In what follows we marginalise over these coefficients in our forecasts. We model the power spectrum of the velocity as in Ref.~\citep{Munoz:2019fkt} using the form we defined in Eq.~\eqref{eq:velo_pert}. We use \texttt{21cmvFAST} to calculate the window function and the amplitude $A_{\rm vel}(z)$ for a given feedback model, and calculate $\Delta_{v^2}^2$ for a given cosmology. We calculate the relative-velocity transfer function at the end of recombination using \texttt{CLASS} Boltzmann solver~\citep{2011JCAP...07..034B}, which we use as an initial condition for our simulations. We show the effect of relative velocities on the 21-cm power spectrum at redshift $z=14$ in Fig.~\ref{fig:display_CIPs_VAOs} for the medium baryonic feedback scenario defined in \texttt{21cmFAST}.

\section{ULAs and their effects on observables}\label{sec:ULAs}

\subsection{ULA equations of motion and the density power-spectra}

A coherently oscillating scalar field $\phi$ in a quadratic potential $V=m_\alpha^2\phi^2/2$ has an energy density that scales as $a^{-3}$ and thus can behave in cosmology as DM~\cite{Turner:1983he}. The Klein-Gordon equation is
\be
(\Box-m_\alpha^2)\phi=0\,,
\ee
where $\Box$ is the D'Alembertian for the cosmological spacetime, and $m_\alpha$ is the axion mass. At zeroth order in cosmological perturbation theory, the solution of this equation for the field initially displaced a fixed amount from the vacuum leads to production of the ULA DM relic density via realignment~\cite{Abbott:1982af,Dine:1982ah,Preskill:1982cy,Turner:1983he}. At higher orders in perturbation theory, the ULA field is coupled to the gravitational potential, and undergoes structure formation. The difference between ULAs and CDM arises due to the gradient terms in $\Box$, which in linear perturbation theory lead to a sound speed in the effective ULA fluid equations~\cite{Khlopov:1985jw,Hu:1998kj,Hu:2000ke,Hwang:2009js}.

In the moving background of the DM-baryon relative velocity the ULA and baryon fluid equations are~\cite{Marsh:2015daa}
\be
\dot{\delta}_\alpha\,&&=\frac{i\mu}{a}\,\delta_\alpha\!-\!\theta_\alpha \,, \label{eqn:dot_delta_a_vbc} \\
\dot{\theta}_\alpha\,&&=\frac{i\mu}{a}\,\theta_\alpha\!-\!\frac{3}{2}H^2[\Omega_\alpha(t)\delta_\alpha\!+\!\Omega_b(t)\delta_b]-2H\theta_\alpha\!+\!\frac{k^2 c_{s,a}^2}{a^2}\delta_\alpha\,, \non \\
\dot{\delta}_b\,&&=-\theta_b \,,  \\
\dot{\theta}_b\,&&=-\frac{3}{2}H^2[\Omega_\alpha(t)\delta_\alpha\!+\!\Omega_b(t)\delta_b]\!-\!2H\theta_b\!+\!\frac{c_{s,b}^2k^2}{a^2}(\delta_b+\delta_T) \,, \non \\ 
\dot{\delta}_T\,&&= \frac{2}{3}\dot{\delta}_b \!+\!\frac{T_\gamma}{T_g}\frac{\Gamma_C}{H(1+z)}\delta_T 
\label{eqn:dot_theta_b_vbc},
\ee
where $\mu=\vec{v}_{b\alpha}^{\,\rm (bg)}(t)\cdot\vec{k}$, $\Omega_\alpha$ is the ULA density, $\Omega_b$ is the baryon density, $\vec{v}_{b\alpha}^{\,\rm (bg)}$ is the ULA-baryon relative velocity, $c_{s,b}$ is the baryon sound-speed, and $c_{s,a}$ is the ULA sound-speed, $\Gamma_C$ is Compton interaction rate, $T_\gamma$ is the CMB temperature, $T_g$ is the gas temperature and we include the effect of temperature fluctuations $\delta_T=\delta_{T_g}/T_g$. On sub-horizon scales, the ULA sound speed is approximated as
\be
c_{s,\alpha}^2\approx \frac{k^2}{4m_\alpha^2 a^2}\, .
\ee

We solve these equations given the initial conditions for a $\Lambda$CDM Universe at {matter-radiation equality that} we compute using CLASS~\citep{2011JCAP...07..034B}. The ULA sound speed leads to a suppression of power relative CDM (which has $c_s=0$). Using CDM initial conditions is not exactly accurate for ULAs, since there is already some departure from CDM at high redshift in the transfer function (see e.g. Refs.~\cite{Hu:2000ke,Hlozek:2014lca}), however the dynamical suppression of structure for $z<1010$ caused by the sound speed gives an accurate model at low $z$ (such an approach was taken in Refs.~\cite{Schive:2014dra,Veltmaat:2016rxo}).

We demonstrate this suppression in clustering statistics with the power-spectra of the baryon and matter fluctuations in Fig.~\ref{fig:pedagogy_1}. In these figures we also show the suppression in clustering due to DM-baryon relative-velocity which increases the baryon pressure experienced by halos. {While on large scales ($k\sim0.1\text{Mpc}^{-1}$) the fluctuations {of the 21-cm temperature are enhanced due to the additional VAO term} (as discussed in Sec.~\ref{sec:impact_of_v}), small-scale fluctuations of {density} are somewhat suppressed by the relative velocities.}

\subsection{VAO amplitude}\label{sec:VAO_amplitude}

\begin{figure*}[t!]
    \vspace{-0.35cm}
    \includegraphics[width=2\columnwidth]{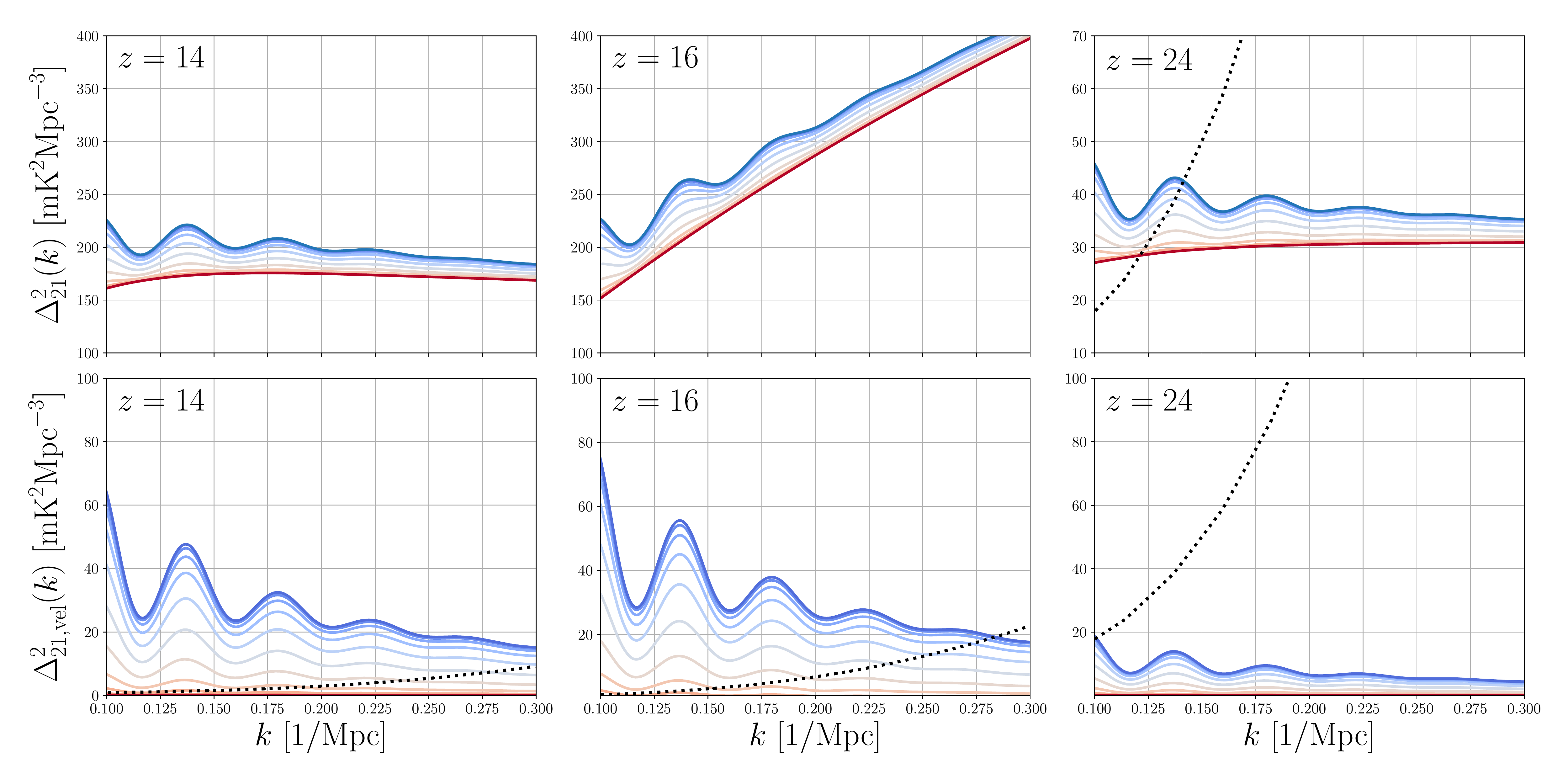}
    \vspace{-0.4cm}
    \caption{\textbf{The effect of ULAs on the VAO signature in the 21-cm power-spectra for redshift bins centred at} $\boldsymbol{z=\{14,16,24\}}$. The VAO amplitude can be shown to vary with $m_\alpha$. The dotted black lines correspond to the anticipated noise (cosmic variance and thermal noise combined) for the HERA experiment, which we calculate with \texttt{21cmSense}. We discuss the binning in wavenumber $k$ in Sec.~\ref{sec:forecasts}. Upper plots correspond the 21-cm signal. The blue end of the color spectrum corresponds to the prediction from the $\Lambda$CDM cosmology, averaged over 20 simulations with \texttt{21cmvFAST}. We demonstrate the effect of ULAs on the 21-cm power spectrum with varying colors. See Fig.~\ref{fig:pedagogy_1} for the correspondence between different colors and the ULA mass. Here, the effect of ULAs are calculated analytically using~Refs.~\citep{2010JCAP...11..007D,Ali-Haimoud:2013hpa,Munoz:2018jwq} as discussed in Sec.~\ref{sec:VAO_amplitude}.}     \label{fig:results_1}
    \vspace{-0.1cm}
\end{figure*}

The amplitude of the VAO signature $A_{\rm vel}(z)$ depends non-linearly on the small-scale brightness temperature, which is sensitive to the baryonic feedback processes and non-linear clustering on smaller scales. Similar to Ref.~\citep{Munoz:2019rhi}, we can calculate the
amplitude of the VAOs by defining a 21-cm relative-velocity bias in the form 
\be
b_{21}^{2}(z)=\langle T_{21}^2({v_{\rm cb}})\rangle-\langle T_{21}({v_{\rm cb})}\rangle ^2
\ee
such that the relative-velocity contribution to the 21-cm power-spectra can then be written as 
\be\label{eq:velo_pert_bias}
\Delta_{21,\rm vel}^2(k,z)=b_{21}^{2}(z)\Delta_{v^2}^2(k,z)|W(k,z)|^2\,,
\ee
and the relative-velocity bias $b_{21}(z)$ has the fiducial value of $A_{\rm vel}(z)$. In what follows, we calculate the derivative of the bias term with respect to the axion mass $\partial b^2_{21}(z)/\partial m_\alpha$ that appears in our forecasts using the analytical approximations up to quadratic order in density fluctuations as described in~\citep{2010JCAP...11..007D,Ali-Haimoud:2013hpa,Munoz:2018jwq}. We set the fiducial value to $A_{\rm vel}(z)$, which we calculate from fitting Eqs.~\eqref{eq:velo_pert}~and~\eqref{eq:smooth_pw} to our simulations using \texttt{21cmvFAST} with no ULAs (equivalent to $m_\alpha\rightarrow \infty$). We demonstrate the effect of ULAs on the 21-cm power-spectra in Fig.~\ref{fig:results_1} {which lead a suppression of the quadratic contribution to the VAO amplitude. As shown in Ref.~\citep{Ali-Haimoud:2013hpa}, because the VAO signature in the 21-cm temperature fluctuations is inherently quadratic, the VAO signature in the 21-cm temperature \textit{power-spectrum} amplitude $\Delta_{21,\rm vel}^2(k)$, at wavenumber $k$, ends up being sensitive to smaller scales than $k$. This allows the VAO signature to be sensitive to small scales where ULAs suppressed the density fluctuations. ULAs cut-off the power spectrum on small scales, and remove the (higher order) modulating effect of the DM-baryon relative velocity on the small scale matter power spectrum. Thus, a cut-off in the power spectrum on $k$ smaller than the baryon smoothing scale drastically reduces the quadratic contribution to the VAO amplitude, as anticipated in Ref.~\cite{Marsh:2015daa}. This is the primary physical effect demonstrated explicitly for the first time in the present work (Fig.~\ref{fig:results_1}), and drives the forecasted constraints that follow.}

Note that the 21-cm temperature-brightness signal at redshift $z\sim$16 corresponds to the halfway point throughout the heating transition marked by large fluctuations compared to the later times ($z\sim$14) where $X$-rays heat up the IGM, making the power-spectrum (and the VAO signature) smaller and less pronounced, as well as earlier times where Lyman-$\alpha$ becomes comparable to the $X$-ray, as demonstrated {by the redshift evolution of the signal visible} in Fig.~\ref{fig:results_1}.

{In a nut-shell, the VAO feature on large scales (the velocity coherence scale) are induced by the second order modulation of the power spectrum by relative velocity on small scales (near the baryon Jeans scale). The amplitude and shape of the VAO decouple, and the amplitude can be estimated by computing the velocity bias, $b_{21}(z)$, in moving background perturbation theory. ULAs damp the power on scales below the axion Jeans scale. Thus, if the axion Jeans scale is larger than the baryon Jeans scale, relative velocities induce no power spectrum modulation in moving background pertrubation theory~\cite{Marsh:2015daa}, the velocity bias vanishes, and thus too do the large scale VAO.}

The analytic approximation to $b_{21}(z)$ has been shown to match the qualitative features of the relative-velocity contribution to the power-spectra while also predicting $A_{\rm vel}(z)$ to be an $\mathcal{O}(1)$ lower compared to more robust calculations using \texttt{21cmvFAST} simulations~\citep{Munoz:2019rhi}, which capture the baryonic feedback processes and the effect of non-linear structure formation more accurately. In what follows, we marginalize over the baryonic feedback strength using our simulations to account for the former, and leave evaluating the effect of non-linear structure formation on the partial derivative $\partial b^2_{21}(z)/\partial m_\alpha$ to future works. Note nevertheless that our fiducial values for $A_{\rm vel}(z)$ are robust as they are calculated from simulations.

\section{Forecasts}
\label{sec:forecasts}

\subsection{Experimental specifications}

We take four redshift bins of size $\Delta z=2$, centred at redshifts $z\in\{12,14,16,18\}$, and one redshift of size $\Delta z=3$, centred at $z=24$ to match Ref.~\citep{Munoz:2019rhi}. These choices are made to get sufficiently high SNR per redshift bin in order to detect the VAO features using experimental specifications matching HERA survey~\citep{DeBoer:2016tnn}.

Foreground mitigation for the cosmological 21-cm signal is performed either via wedge suppression or avoidance\footnote{Please see \cite{2014PhRvD..90b3018L, 2014PhRvD..90b3019L} for detailed description of the EoR window and foreground wedge.}.
\texttt{21cmSense} \citep{Pober:2012zz, Pober:2013jna,  2016ascl.soft09013P} is a python module designed to estimate the noise power spectra when a given telescope array observe the 21-cm signal via foreground avoidance. We use \texttt{21cmSense} for the HERA experiment, which we describe below.

In every $uv$ bin\footnote{The $uv$ plane is the Fourier transform of the sky brightness distribution in a plane perpendicular to the direction of observation. Radio interferometers cannot produce an image of the sky directly, instead they make observations in the $uv$ plane.} the noise is calculated as,
\begin{equation}\label{eq: 21cmsense_uv_bins}
\delta^2_{\rm uv}(\boldsymbol{k}) \approx X^2Y ({\boldsymbol{k}^3}/{2 \pi ^2}) ({\Omega_{\rm Eff}}/({2t_{0}})) T_{\rm sys}^2,
\end{equation}
where $X^2Y$ is a scalar conversion from an observed solid angle (or effective beam, $\Omega_{\rm Eff}$) to a comoving distance \cite{2014ApJ...788..106P}, $T_{\rm sys}$ is the system temperature, $t_0$ is the total observation time and $\boldsymbol{k}$ is the three-dimensional Fourier wavenumber. 

Assuming Gaussian errors on cosmic variance, we express the total uncertainty with an inversely-weighted sum across all the $k$ modes as
\begin{equation}\label{eq: Tscope_noise}
\Delta_{21}^{2,\rm obs}(k) = \left\{ \sum_i\frac{1}{[\delta^2_{{\rm uv},i}(k)+\Delta^2_{21}(k)]^2}\right\}^{-\frac{1}{2}} ,
\end{equation}
where the index, $i$, represents multiple measurements of the same frequency from redundant baselines within the array.
This is therefore the total noise, including both sample variance and thermal noise.

The foreground wedge is defined as 
\be\label{eq: wedge}
k_\parallel = a + b k_\perp\,,
\ee
where $k_\parallel$ and $k_\perp$ are the Fourier modes projected on the line-of-side and the transverse plane respectively; $b$ depends on the instrument beam, bandwidth and underlying cosmology; 
$a$ is the buffer zone, typically chosen as $a=0.1h\text{Mpc}^{-1}$ ($a=0.01h\text{Mpc}^{-1}$) for pessimistic (optimistic) scenarios.

We apply \texttt{21cmSense} to HERA \cite{2016ApJ...826..181D, 2015ApJ...800..128B}, 
where stations are located in a filled hexagonal grid (11 along each side).
Each station is 14 $\rm m$ in diameter giving a total collecting area of 50,953 $\rm m^2$ accross a total bandwidth ranging $[50, 250]~ \rm MHz$.
The antennae are taken to be at $T_{\rm rx} = 100 \rm K$.
HERA is operated only in drift scan mode for 6 hours per night. Throughout the paper we assume three years of observation with HERA.

For reference, we calculate the total SNR as equal to $\sum_k\Delta^2_{21}(k,z)/\Delta_{21}^{2,\rm obs}(k,z)$ and the VAO SNR as equal to $\sum_k\Delta^2_{21,\rm vel}(k,z)/\Delta_{21}^{2,\rm obs}(k,z)$, summed (in  quadrature) over the redshift bins we consider. Here, $\sum_k$ is the sum over binned wavenumbers and $\Delta_{21}^{2,\rm obs}(k,z)$ is the total uncertainly of the 21-cm spectra defined in Eq.~(13). We set the $k$-bin widths to the inverse of the cosmological bandwidth corresponding to the redshift range that can be considered co-eval as in~\citep{DeBoer:2016tnn}. We find the total (over all $z$) detection signal-to-noise ratio (SNR) of the 21-cm signal to be $\{130,190,498\}$, and the SNR of the VAO signature to be $\{21,40,190\}$, for our $\{$pessimistic,\,moderate,\,optimistic$\}$ foreground considerations, respectively, and for regular baryonic feedback, using HERA~\citep{DeBoer:2016tnn}.

\subsection{Results}

We define the information matrix as 
\begin{equation}
\label{eq:fisherM}
F_{\alpha\beta}(z) = \sum\limits_{k-\rm bins}\frac{\partial_\alpha\Delta_{21}^2(k,z)\partial_\beta\Delta_{21}^2(k,z)}{(\Delta_{21,\rm obs}^{2})^2(k,z)}\,,
\end{equation}
and model the 21-cm power-spectra with 
\be\label{eq:parameters}
\{c_0(z),c_1(z),c_2(z),c_3(z),c_4(z),m_\alpha,\epsilon_{\rm FB}(z)\}\,,
\ee
at each redshift bin we consider, where $c_i(z)$ with $i\in\{0,\ldots,4\}$ parametrises the non-wiggle part of the 21-cm power-spectra, $m_\alpha$ is the ULA mass, and $\epsilon_{\rm FB}$ is a parameter representing the sensitivity of the 21-cm power-spectra {to the Lyman-Werner (LW) baryonic feedback efficiency in the \texttt{21cmFAST} simulations we consider, which parametrise the formation of stars as a change of mass of cooling haloes as~\citep{2013MNRAS.432.2909F,Visbal:2014fta}
\begin{equation}
M_{\rm cool}(z,v_{\rm cb},F_{\rm LW})=M_{\rm cool}(z,v_{\rm cb},0)\times[1+\epsilon_{\rm FB}B(F_{\rm LW})^{\beta}]\,,
\end{equation}
where $M_{\rm cool}(z,v_{\rm cb},0)$ is the mass of cooling in the absence of LW feedback, $F_{\rm LW}=1.256\times10^{-20}{\rm ergs}\,{s}^{-1}{\rm cm}^{-2}{\rm Hz}^{-1}{\rm sr}^{-1}$ is the LW flux, $B=4$ and $\beta=0.47$ as fitted to the simulations in Refs.~\citep{Machacek:2000us,Wise:2007nb}.}

\begin{figure}[t!]
    \includegraphics[width=\columnwidth]{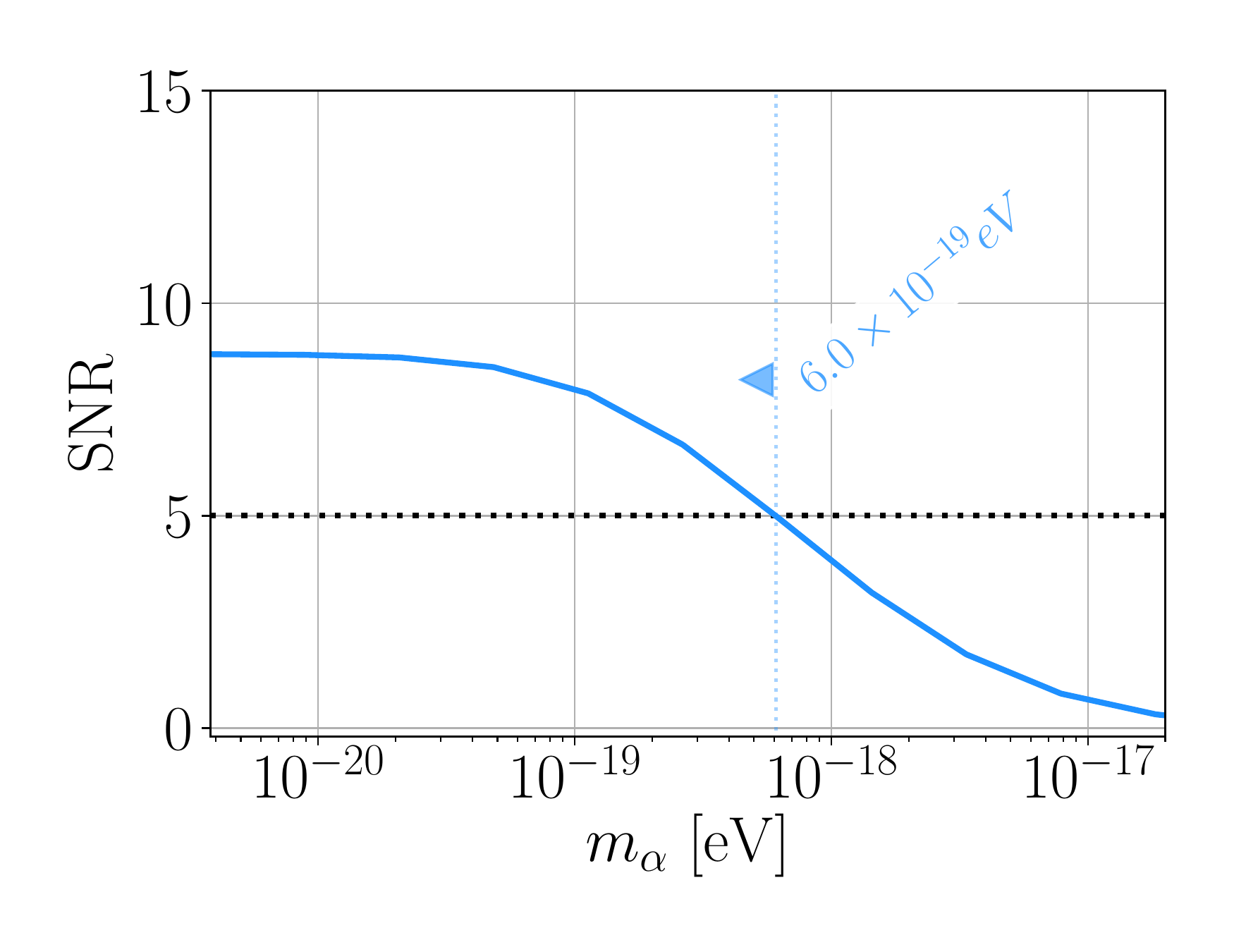}
    \vspace{-1.2cm}
    \caption{The total SNR on the ULA mass using foreground cut-off scale {$a=0.05h\text{Mpc}^{-1}$}, assuming moderate baryonic feedback and HERA survey specifications as we describe in the text. The SNR of the ULA mass detection is plotted for varying ULA mass $m_\alpha$. The vertical lines indicate that ULA masses lower than {$m_\alpha=6.0 \times10^{-19}\, \text{eV}$} will be detected with an SNR of {5} or higher. The horizontal lines correspond to SNR of 3. The SNR goes to zero for high ULA masses and flattens at low masses where scenarios with different ULA masses cannot be distinguished from each other.}
    \vspace{-0.45cm}
    \label{fig:mass_constraints_different_cuts4new_fg}
\end{figure}

In Fig.~\ref{fig:mass_constraints_different_cuts4new_fg}, we show the total signal-to-noise (SNR) on the ULA mass from the five redshift bins defined above. The colored vertical lines indicate that ULA masses lower than {$m_\alpha=6.0\times10^{-19}\,\text{eV}$} will be detected with an SNR of {5} or higher. The dotted black horizontal line correspond to SNR of {5}. The SNR goes to zero for high ULA mass where the effect of the ULA on the 21-cm is negligible. The SNR flattens at low masses where models with different ULA masses cannot be distinguished from each other. Here, the buffer zone is set as $a=0.05h\text{Mpc}^{-1}$, corresponding to our \textit{moderate} foreground scenario we describe next.

\begin{figure}[t!]
    \includegraphics[width=\columnwidth]{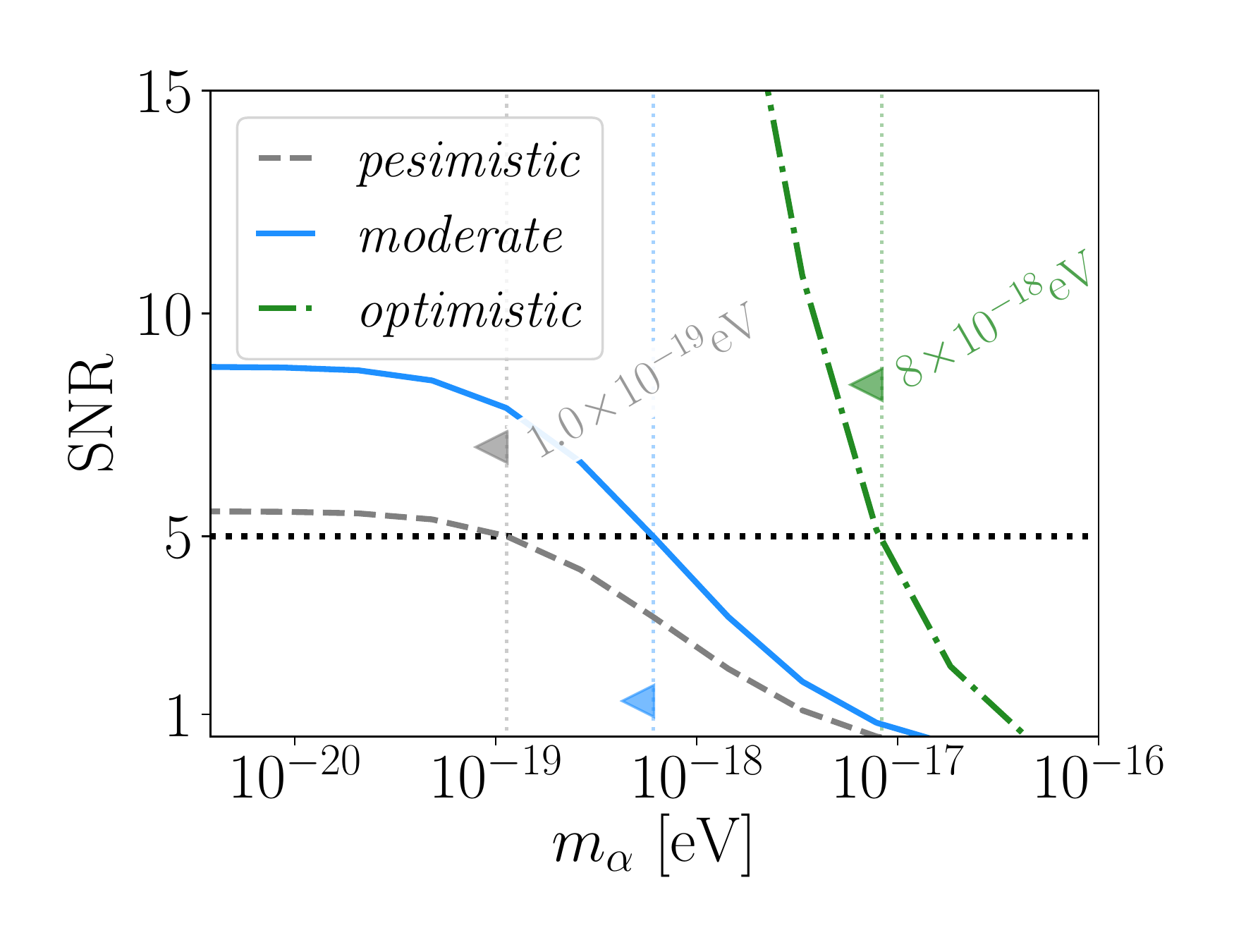}\\
    \vspace{-0.5cm}
    \vspace{-0.5cm}
    \caption{Analysis of the effect of foregrounds on the ULA mass constraints. Three foreground scenarios are shown, these are \textit{pessimistic} (dashed grey line), moderate (solid blue line) and \textit{optimistic} (dot-dashed green line). The upper limits on the ULA mass can be seen to worsen for the \textit{pessimistic} foreground scenario where masses below {$1.0\!\times\!10^{-19}{\rm eV}$} are constrained with {${\rm SNR}>5$}. For the \textit{optimistic} scenario, ULA masses below {8.0}$\times10^{-18}\,\text{eV}$ are constrained with SNR above 5. The \textit{moderate} scenario is identical to the results shown if Fig.~\ref{fig:mass_constraints_different_cuts4new_fg}.}
    \vspace{-0.45cm}
    \label{fig:mass_constraints_different_cuts5new_fg}
\end{figure}

Foregrounds play an important role in determining our ability to infer cosmology from the 21-cm signal. We demonstrate how various foreground scenarios affect the constraints on the ULA mass in Fig.~\ref{fig:mass_constraints_different_cuts5new_fg}. Here, we have defined the scenario where the baselines are added incoherently, no $k$ modes are included from within the horizon wedge (and buffer zone) and $a=0.1h\text{Mpc}^{-1}$ as \textit{pessimistic}. In the \textit{moderate} scenario, the baselines are added coherently and $a=${0.05}$h\text{Mpc}^{-1}$, otherwise same as the \textit{pessimistic} scenario. In the \textit{optimistic} scenario, all baselines in the primary field of view (no buffer zone) are added coherently and $a=${0.03}$h\text{Mpc}^{-1}$. We find ULA masses below {$1.0\!\times\!10^{-19}\,\text{eV}$} and {$8\!\times\!10^{-18}\,\text{eV}$} will be detected to SNR of 3 or higher for the pessimistic and optimistic scenarios, respectively. The SNR {exceeds 10} by {$m_\alpha=10^{-18}\,\text{eV}$} in the optimistic scenario, while flattening {around} $\sim5$ in the pessimistic scenario.

Another important factor when detecting the ULAs is the effect of baryonic feedback, which suppresses the VAO signature~\citep{Munoz:2019rhi}. Moreover, if feedback affects the 21-cm in a similar way, we might expect the ULA mass to be somewhat degenerate with $\epsilon_{\rm FB}$. Indeed in Fig.~6 we show that for a single redshift bin, the effect of baryonic feedback has some degeneracy with $m_\alpha$. Here, we show the $m_\alpha-\epsilon_{\rm FB}$ contour plots from various combinations of the redshift bins. The foremost solid contour in indigo corresponds to the constraints from the redshift bin centered at $z=14$. Following dashed contours corresponds to successively adding information from redshift bins centered at $z=\{24,18,16,12\}$, with colors described in the figure caption. Innermost blue solid contour correspond to the forecasts from adding all redshift bins, matching our \textit{moderate} foreground scenario. The redshift information can be seen to improve the ULA mass estimate along the degeneracy direction suggesting the degeneracy between the ULA mass and the feedback parameter $\epsilon_{\rm FB}$ at $z=14$ is somewhat broken. 

Lastly, we expect our forecasts to depend on assumptions about baryonic feedback, which affects the large-scale 21-cm spectrum by preventing smaller-mass molecular-cooling haloes to form stars, in turn the reducing the effect of relative-velocities on the 21-cm fluctuations. In Fig.~\ref{fig:mass_constraints_different_cuts8new_fg}, we show the ULA mass constraints for different feedback models defined in \texttt{21cmFAST}. Indeed, lower (higher) feedback levels lead to more optimistic (pessimistic) constraints.

\begin{figure}[t!]
    \includegraphics[width=0.95\columnwidth]{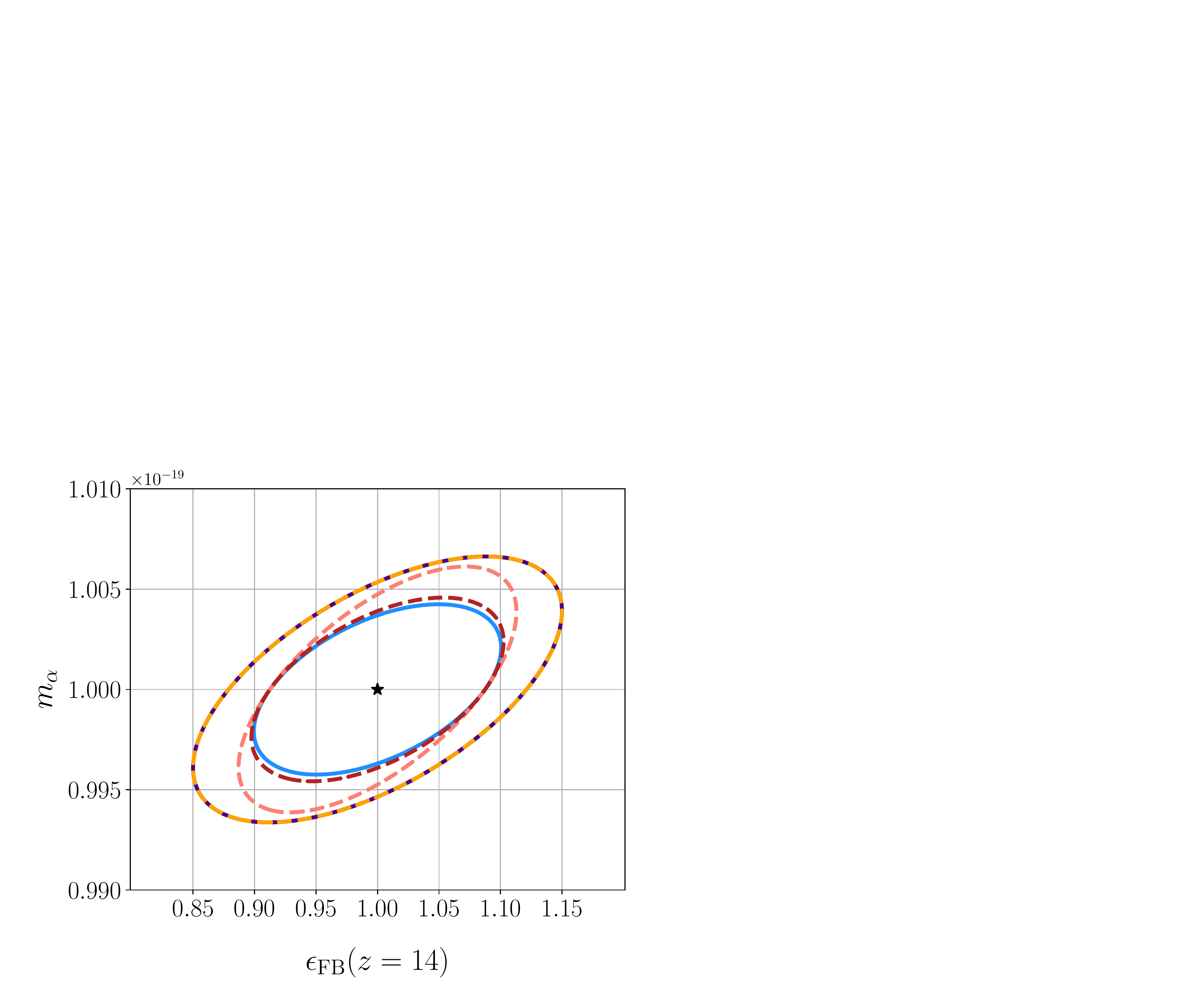}
    \vspace{-0.3cm}
    \label{fig:results_fisher}
    \begin{center}
        \begin{tabular}{l @{\hskip 12pt} c@{\hskip 12pt}c@{\hskip 12pt}c@{\hskip 12pt}c} 
            \toprule
           Description of the observables  &   \multicolumn{2}{c}{Color} \\ [0.5ex] 
           \hline
            $z_{\rm bin}\in\{14\}$ & \textcolor{indigo}{\rule{1cm}{0.75mm}} & Indigo \\ 
            $z_{\rm bin}\in\{14, 24\}$ & \textcolor{orange}{\rule{1cm}{0.75mm}} & Orange \\
            $z_{\rm bin}\in\{14, 18, 24\}$ & \textcolor{salmon}{\rule{1cm}{0.75mm}} & Salmon \\ 
            $z_{\rm bin}\in\{14,16, 18,24\}$ & \textcolor{purple}{\rule{1cm}{0.75mm}} & Brown \\
            $z_{\rm bin}\in\{12, 14, 16, 18, 24\}$  & \textcolor{blue}{\rule{1cm}{0.75mm}} & Blue \\
           \hline
        \end{tabular}
   \end{center}
    \vspace{-0.2cm}
    \caption{Contour plots demonstrating the improvement on the ULA mass constraints from the redshift information, {for fiducial values of $m_{\alpha}=10^{-19}{\rm eV}$ and $\epsilon_{\rm FB}=1$.} The foremost contour (solid, indigo colour) corresponds to the constraints from a single redshift bin centered at $z=14$. Following dashed contours correspond to successively adding information from redshift bins centered at $z=\{24,18,16,12\}$ with colors described above. Innermost plot correspond to adding all redshift bins, matching our \textit{moderate} foreground scenario (shown with blue solid contour) forecasts described before. The redshift information can be seen to improve the ULA mass estimate along the degeneracy direction. {We find the VAO measurements can potentially measure the axion mass of $m_{\alpha}=10^{-19}{\rm eV}$ over 95\% precision.}}
\end{figure}
\begin{figure}[t!]
    \includegraphics[width=\columnwidth]{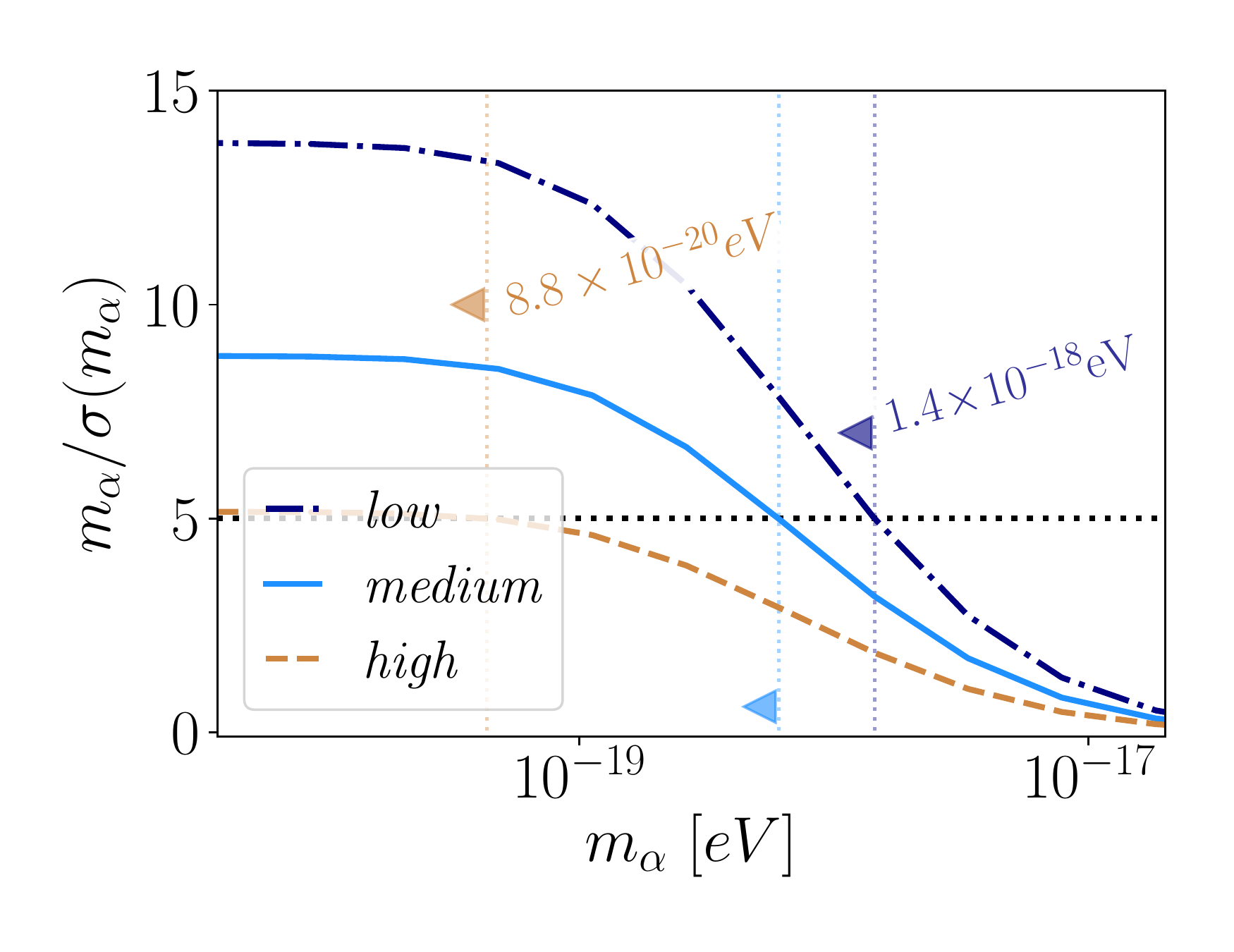}
    \vspace{-0.7cm}
    \caption{Analysis of baryonic feedback assuming three feedback levels as defined in \texttt{21cmFAST}. Higher baryonic feedback strength prevents the forming of smaller-mass halos reducing the effect of relative-velocities on the 21-cm spectra on large scales. The blue solid line correspond to the \textit{medium} feedback scenario assumed throughout the rest of this paper. Dot-dashed dark blue (dashed brown) line correspond to \textit{low} (\textit{high}) feedback scenario. Similar to before, we show the 3 sigma constraints on the ULA mass with vertical lines matching the different lines in color.}
    \vspace{-0.45cm}
    \label{fig:mass_constraints_different_cuts8new_fg}
\end{figure}

\section{Discussion and Conclusions}\label{sec:discussion}

Ultra-light axions (ULAs) are a well-motivated candidate to compose all or some of the dark matter and can have a wide range of particle masses. The velocity acoustic oscillation (VAO) signature in the 21-cm spectrum on large scales provides a window to test ULAs through its sensitivity to small-scale fluctuations in gas density and temperature. This signal is most pronounced during the cosmic dawn era ($z\sim10-30$) where the typical mass of collapsed baryonic objects falls near the critical mass below which gas pressure prevents their collapse. 

We have evaluated the detection prospects of ULAs with measurements of the 21-cm fluctuations from cosmic dawn. The ongoing HERA experiment may be able to constrain ULA mass below {$6.0\times10^{-19}\,\text{eV}$ at SNR of 5} for moderate foreground and feedback scenarios. The sensitivity weakens by a factor $\sim4$ if the foregrounds end up more detrimental or by $\sim3$ if the baryonic feedback strength is larger. Conversely, if the foregrounds are lower, the sensitivity can be higher by a factor of $\sim5$. Similarly, lower baryonic feedback can also improve the ULA sensitivity by a factor of $\sim2$. Improvements such as this bridge the gap between large scale existing cosmological bounds, and small-scale constraints from galaxy evolution and black-hole physics~\cite{Arvanitaki:2010sy,Marsh:2018zyw,Stott:2018opm}, closing the gaps in ULA parameter space~\cite{Grin:2019mub}. 

Much is yet unknown about the astrophysics of the cosmic dawn era. In particular, our ability to detect the VAO signature to high significance to infer cosmology depends on the baryonic feedback strength and the severity of foreground contamination. Here, we have shown how these complications affect constraints to ULAs using a simple information-matrix analysis and assuming phenomenological parameters represent the underlying cosmology and astrophysics. Going forward, we could build upon the observations made in this paper by more rigorous simulations with wider-ranging assumptions and modelling to test our predictions. Nevertheless, we foresee a bright future for the cosmological significance of the cosmic-dawn signal, and its potential role in constraining ULA mass.

\section{Acknowledgements}
 SCH thanks T. Binnie and J.~B.~Mu$\tilde{\rm n}$oz for useful conversations.  SCH is supported by the Horizon Fellowship from Johns Hopkins University. DJEM is supported by an Ernest Rutherford Fellowship from the Science and Technologies Facilities council of the United Kingdom. This work was performed in part at Aspen Center for Physics, which is supported by National Science Foundation grant PHY-1607611, and a grant from the Simons Foundation.  This work was supported at JHU in part by the Simons Foundation and by National Science Foundation grant No.\ 2112699.

\bibliography{vao_constraints.bib}

\end{document}